\documentclass[10pt,aps,prd,amsmath,amssymb,superscriptaddress,twocolumn,nofootinbib,showpacs,preprintnumbers,amsfonts, notitlepage]{revtex4-2}
\usepackage{cancel}
\usepackage[usenames,dvipsnames,svgnames,table]{xcolor} 
\usepackage{graphicx}
\usepackage{dcolumn}
\usepackage{bm}
\usepackage{hyperref}
\hypersetup{ 
    pdfnewwindow=true,      
    colorlinks=true,       
    allcolors=[RGB]{31 119 180}
} 
\usepackage{cleveref}
\usepackage{dsfont}
\usepackage{placeins}
\usepackage{todonotes}
\usepackage[utf8]{inputenc}
\usepackage{gensymb}
\usepackage{multirow}
\usepackage{enumitem}
\usepackage{array}   
\newcolumntype{L}{>{$}l<{$}} 
\newcolumntype{R}{>{$}r<{$}}
\newcolumntype{C}{>{$}c<{$}}
\usepackage{xspace}
\usepackage{braket}
\usepackage{units}
\usepackage{slashed}
\usepackage[normalem]{ulem}
\usepackage[format=plain,justification=RaggedRight,singlelinecheck=false]{caption}
\usepackage[format=plain,justification=centering,singlelinecheck=false]{subcaption}

\newcommand{\mytitle}[1]{\vspace{.5cm}{\em #1.---}}
\newcommand{\Pc}{$P_c(4312)$}
\newcommand{\SigmaD}{$\Sigma_c^+\bar{D}^0$}
\newcommand{\abs}[1]{\ensuremath{|#1|}}

\let\Im\relax

\DeclareMathOperator{\Im}{Im}

\DeclareMathOperator{\sign}{Sign}

\newcommand{\eg}{{\it e.g.}\xspace}

\newcommand{\ie}{{\it i.e.}\xspace}

\newcommand{\mevnospace}{\ensuremath{{\mathrm{\,Me\kern -0.1em V}}}}
\newcommand{\gevnospace}{\ensuremath{{\mathrm{\,Ge\kern -0.1em V}}}}
\newcommand{\tevnospace}{\ensuremath{{\mathrm{\,Te\kern -0.1em V}}}}
\newcommand{\mev}{\mevnospace\xspace}
\newcommand{\gev}{\gevnospace\xspace}

\newcommand{\jpsi}{\ensuremath{J/\psi}\xspace}

\newcommand{\jpsip}{\ensuremath{J/\psi\,p}\xspace}

\newcommand{\catania}{INFN Sezione di Catania, Catania, I-95123, Italy}
\newcommand{\ceem}{Center for  Exploration  of  Energy  and  Matter,
Indiana  University,
Bloomington,  IN  47403,  USA}
\newcommand{\epfl}{Institute of Physics, Ecole Polytechnique F\'ed\'erale de Lausanne (EPFL), CH-1015 Lausanne, Switzerland}
\newcommand{\icn}{Instituto de Ciencias Nucleares, 
Universidad Nacional Aut\'onoma de M\'exico, Ciudad de M\'exico 04510, Mexico}
\newcommand{\icsup}{Pedagogical University of Krakow, 30-084 Krak\'ow, Poland}
\newcommand{\indiana}{Department of Physics,
Indiana  University, Bloomington,  IN  47405,  USA}
\newcommand{\fsu}{Department of Physics,
Florida State  University, Tallahassee,  FL  32306,  USA}
\newcommand{\jlab}{Theory Center, Thomas  Jefferson  National  Accelerator  Facility, Newport  News,  VA  23606,  USA}
\newcommand{\rome}{INFN Sezione di Roma, Roma, I-00185, Italy}
\newcommand{\mift}{Dipartimento di Scienze Matematiche e Informatiche, Scienze Fisiche e Scienze della Terra,
Universit\`a degli Studi di Messina, Messina, I-98166, Italy}
\newcommand{\ub}{Departament de F\'isica Qu\`antica i Astrof\'isica and Institut de Ci\`encies del Cosmos, Universitat de Barcelona, Mart\'i i Franqu\`es 1, E08028, Spain}
\newcommand{\ucm}{Departamento de F\'isica Te\'orica, Universidad Complutense de Madrid and IPARCOS, E-28040 Madrid, Spain}
\newcommand{\uned}{Departamento de F\'isica Interdisciplinar, Universidad Nacional de Educaci\'on a Distancia (UNED), Madrid E-28040, Spain}

\begin{document}
\title{Deep Learning Exotic Hadrons}

\preprint{JLAB-THY-21-3518}
\author{L.~Ng}
    \email{ln16@my.fsu.edu}
    \affiliation{\fsu}
\author{\L.~Bibrzycki}
    \email{lukasz.bibrzycki@up.krakow.pl}
    \affiliation{\icsup}
\author{J.~Nys}
\email{jannes.nys@epfl.ch}
    \affiliation{\epfl}
\author{C.~Fern\'andez-Ram\'irez}
    \email{cesar@jlab.org}
    \affiliation{\uned}
    \affiliation{\icn}
\author{A.~Pilloni}
\email{alessandro.pilloni@unime.it}
    \affiliation{\mift}   
    \affiliation{\catania}
    \affiliation{\rome}
\author{V.~Mathieu}
    \affiliation{\ub}
    \affiliation{\ucm}
\author{A.~J.~Rasmusson}
    \affiliation{\indiana}
\author{A.~P.~Szczepaniak}
    \affiliation{\indiana}
    \affiliation{\jlab}
    \affiliation{\ceem}
\collaboration{Joint Physics Analysis Center}
\date{\today}

\begin{abstract}
We perform the first
amplitude analysis
of experimental data
using Deep 
Neural Networks
to determine the nature
of an exotic hadron. Specifically, we study the
line shape of the
$P_c(4312)$ signal reported by the LHCb collaboration and we find that its most likely interpretation 
is that of a virtual state.
This method can be 
applied to
other near-threshold 
resonance candidates. 
\end{abstract}

\maketitle

\mytitle{Introduction}
Many hadron candidates that
deviate from the quark model expectations~\cite{GellMann:1964nj,*Zweig:1981pd}
have been discovered in the last
years~\cite{pdg,Olsen:2017bmm}.
The field of hadron spectroscopy
has flourished
attempting to provide 
a comprehensive picture of the new states.
Many different approaches have been proposed
to explain their underlying nature, 
becoming a playground for testing new
techniques and novel physical
interpretations~\cite{Esposito:2016noz,*ali2019multiquark,Guo:2017jvc,Guo:2019twa,Lebed:2016hpi,*Karliner:2017qhf,*Brambilla:2019esw}.

To determine if an experimental signal corresponds to a 
hadron resonance, it is
necessary to perform an amplitude analysis in order to extract its
physical  properties such as mass, width, couplings,
and quantum numbers.
Most of the data analyses follow
a top-down approach, where the amplitudes
are derived from a microscopic model.
The advantage 
is that it assigns
a physical interpretation to the signal.
The caveat is that the results are 
biased by the assumed dynamics. 
Another possibility is to proceed bottom-up. 
By considering a number of minimally-biased amplitudes compatible with physical principles and
fitting them to data,
one can determine
the existence and properties of resonances in the least model-dependent way. 
Even though in this approach
there is no assumed microscopic model
it is still possible to deduce
the nature of the underlying dynamics
from the analytic properties of the amplitudes.

For example,
both methods were recently used to provide an
interpretation of the
\Pc\ signal
found by LHCb 
in the $\Lambda_b^0 \to K^- \jpsi\, p$ decay~\cite{Aaij:2019vzc}.
This measurement is of particular interest because,
if due to a resonance, it would contain
five valence quarks, which is 
beyond the baryon lore.
The signal peaks approximately $5\mev$ below the \SigmaD\ threshold,
making it a primary candidate for a hadron molecule.
Near-threshold enhancements of the cross section
are known phenomena in particle physics,
\eg the weakly-bound deuteron
in proton-neutron scattering.
The molecular interpretation was found to be compatible with the data
in~\cite{Du:2019pij,*Du:2021fmf}.
Another microscopic interpretation
is that the signal is
a kinematical effect generated by particle
rescattering~\cite{Nakamura:2021qvy}.
The \Pc\ pole position was first obtained 
in~\cite{Fernandez-Ramirez:2019koa} following
a bottom-up approach,
favoring a virtual state interpretation,
\ie an attractive interaction that is not strong enough to bind a state, as it happens, for example, in neutron-neutron scattering~\cite{Hammer:2014rba}.

The evolution of computing capabilities during the last decades
has allowed to develop and employ
powerful numerical 
techniques to unravel
the structure of matter,
with machine learning 
acquiring a prominent role. 
In theoretical hadron physics,
techniques such as 
genetic algorithms~\cite{Ireland:2004kp,*Fernandez-Ramirez:2008ixe,*NNPDF:2014otw}
and neural networks~\cite{Forte:2002fg,*Rojo:2004iq,*Keeble:2019bkv,*Adams:2020aax}
have been exploited as fitters and/or interpolators.
Recently, the idea of using deep 
neural networks (DNN) as model classifiers 
was benchmarked
against the well-known nucleon-nucleon 
bound state~\cite{Sombillo:2020ccg} and pion-nucleon resonances~\cite{Sombillo:2021rxv,*Sombillo:2021yxe}.

In this work we develop and benchmark a systematic approach to
apply DNN as a model-independent tool to analyze
and interpret experimental data.
Following the bottom-up strategy,
we construct generic amplitudes
to train the DNN. 
We then use it as a model classifier 
to infer the physical content of the data.
\color{black}One clear advantage compared to standard $\chi^2$ fits is that DNN determine the probability of each physical interpretation, as they learn the subtle classification boundary between them. 
We teach DNN how to recognize these different phases by targeting the specific regions of the parameter space (which yield stable solutions) that might be difficult to reach during optimization, or might require high resolution spectra to detect.
The result of this is that we no longer need to explore large parameter spaces, but can use DNN to efficiently extract information from the spectra that determines on which side of the  boundary we are located. 
\color{black}
As proof of concept, 
we apply this method to the \Pc\ signal.

\begin{figure}
    \includegraphics[width=\columnwidth]{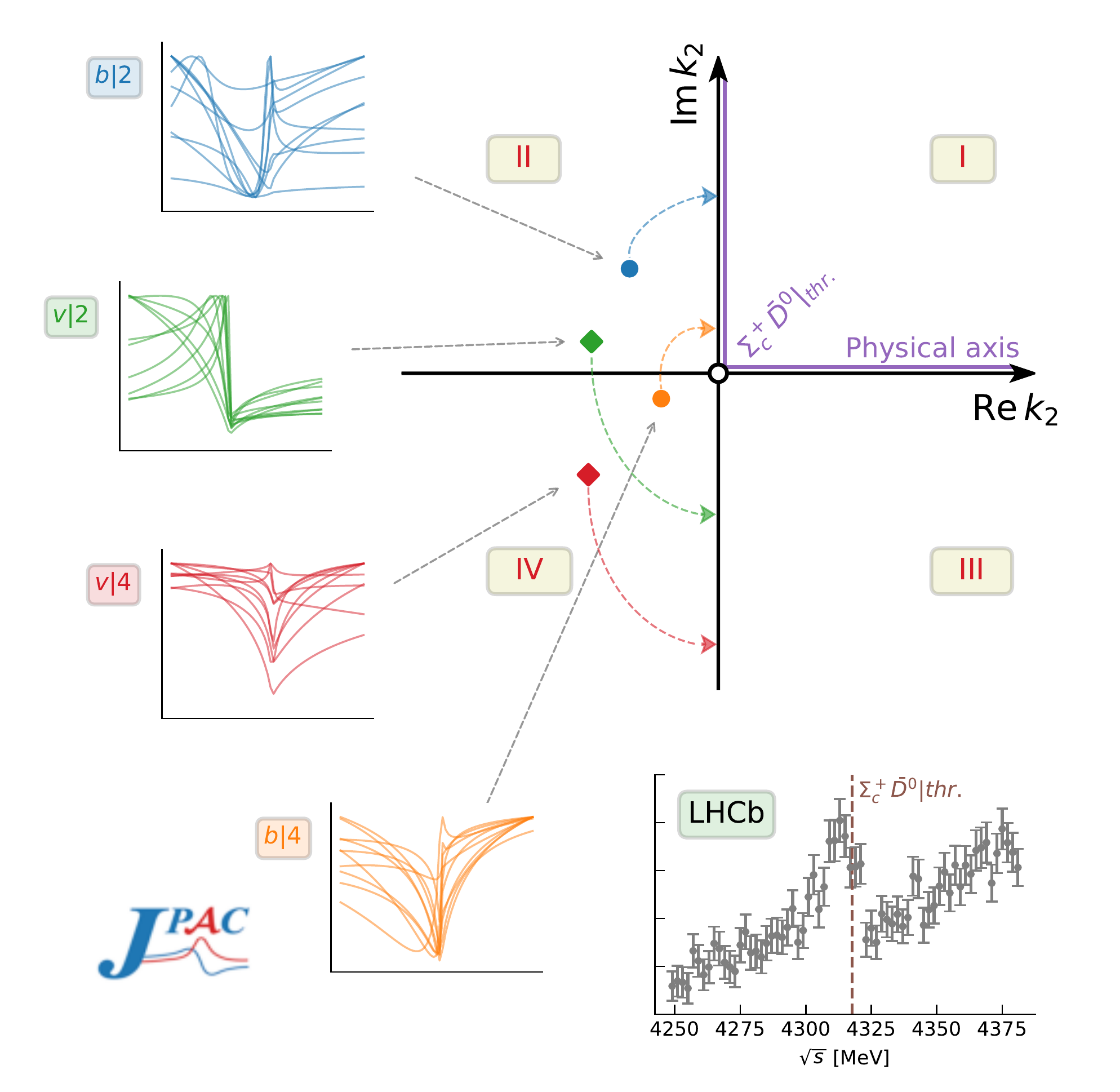} 
      \caption{Analytic structure of the amplitude near the 
      \SigmaD\ threshold. The adjacent Riemann sheets are continuously connected along the axes. The four possible resonant pole structures, that correspond to the classification classes,
      are depicted together with example line shapes. When the \jpsip and \SigmaD\ channels decouple, the poles move to the imaginary $k_2$ axis along paths by the arrows. Poles moving to the positive (negative) axis correspond to bound (virtual) states.
      The bottom-right inset shows the data from LHCb in the \Pc\ region. The layout of the figure is inspired by~\cite{Dudek:2016cru}.
       }\label{fig:complexplane}
\end{figure}

\begin{figure}
    \includegraphics[width=.8\columnwidth]{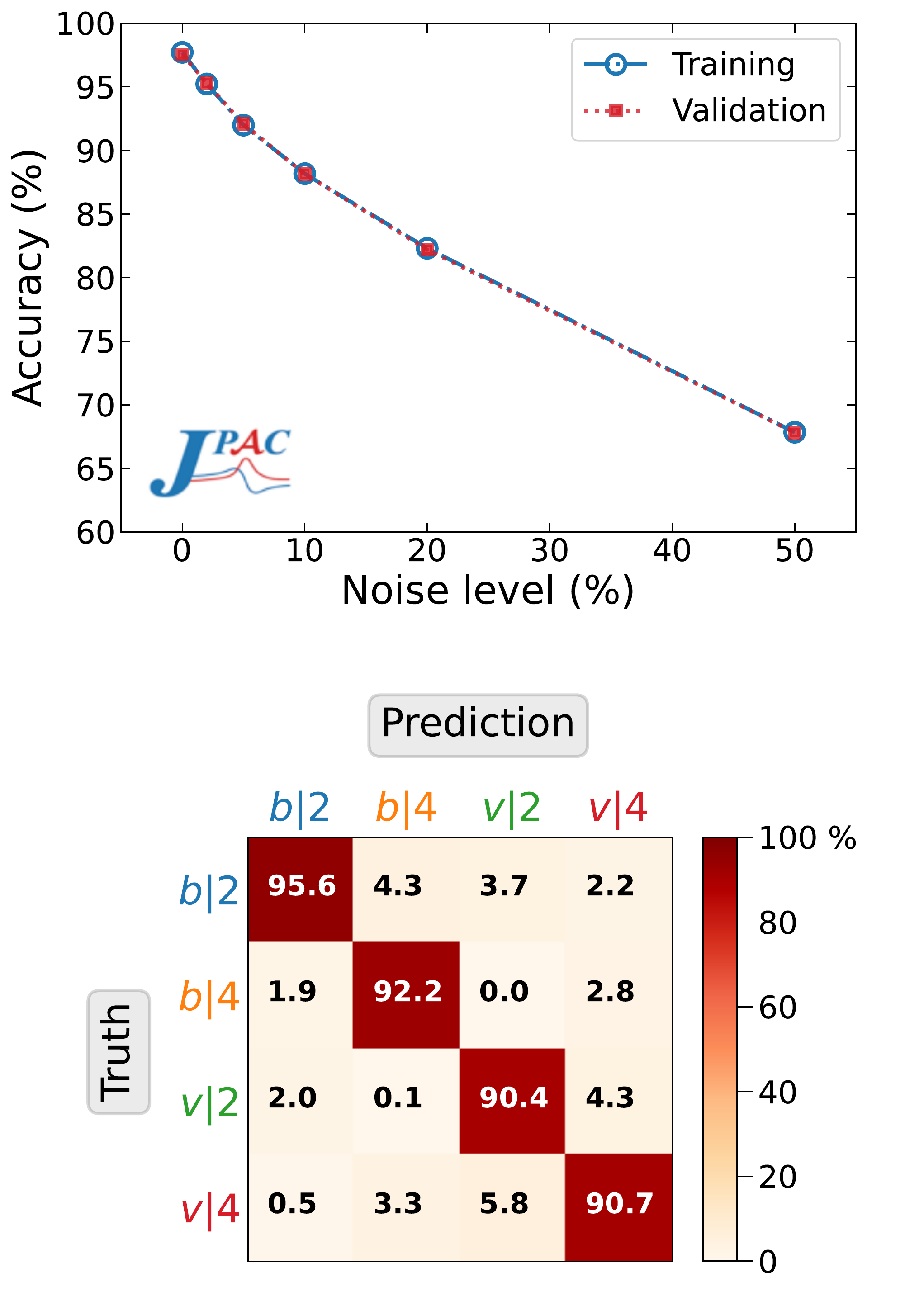}
    \caption{(top panel) Accuracy as a function of the noise level $\sigma$ of the training and validation datasets after 100 epochs in the $\left[4.251,4.379\right]\gev$ energy range. (bottom panel) Confusion matrix for the case of $\sigma=5\%$. Percentages in the confusion matrix refer to the prediction accuracy.}
    \label{fig:confusedaccuracy}
\end{figure}

\mytitle{Physics basis for the neural network}
We focus on the  \jpsip\  invariant mass distribution reported by LHCb in~\cite{Aaij:2019vzc}. This can be parametrized as~\cite{Frazer:1964zz,Fernandez-Ramirez:2019koa}
\begin{equation}
I(s) = \rho(s) \left[|P(s)\, T(s)|^2 + B(s)\right]\, ,
\label{eq:intensity}
\end{equation}
where $s$ is the \jpsip\ 
invariant mass squared,
$B(s)$ and $P(s)$ are smooth functions, and $\rho(s)$ the three-body phase space.
The amplitude $T(s)$ encodes the dynamics of the $J/\psi\,p$ rescattering, and in particular contains the details of the \Pc. Close to the
\SigmaD\ threshold, it can be expanded:
\begin{align}
T(s) &=\frac{m_{22}-ik_2}{(m_{11}-ik_1)(m_{22}-ik_2)-m_{12}^2},
\label{eq.ampl}
\end{align}
where $k_1 = \sqrt{s- (m_\psi+m_p)^2}$ and  $k_2 = \sqrt{s- (m_{\Sigma^+_c}+m_{\bar{D^0}})^2}$.
This Taylor expansion could originate from any microscopic model.
The question is what is the range of validity. If other singularities closeby were present, the expansion would break down in this region. The impact of triangle singularities was estimated to be small  in~\cite{Aaij:2019vzc}, and so was the relevance of higher order terms in the expansion~\cite{Fernandez-Ramirez:2019koa}. It should be noted though, that if other singularities were close enough to impact, the model that describes them could be included in the training set of the DNN. In this case we mean to benchmark the approach by comparing to the known result  in~\cite{Fernandez-Ramirez:2019koa} as described by Eq.~\eqref{eq.ampl}.\color{black}

This function can be analytically continued for complex values of $s$. Since the square roots are multi-valued, the amplitude maps onto four Riemann sheets, represented in Fig.~\ref{fig:complexplane}. 
By construction, this amplitude has four poles in the complex $s$ plane. Two of them are a conjugated pair that appears either on the II or IV sheet, close to the \SigmaD\ threshold where the expansion holds. The other two poles lay far away from the region of interest and have no physical interpretation. The pole position and sheet affect the observed line shape.
Since the Eq.~\eqref{eq:intensity}
is based on an expansion around the
\SigmaD\ threshold,
it is only reliable in its vicinity.
We thus have to ensure 
that the DNN learns
from the appropriate
invariant mass window.

If $m_{12}\to 0$, the \SigmaD\ channel decouples from the \jpsip\ one. In this limit, the $P_c(4312)$ pole would become either a stable bound state, or a virtual threshold enhancement, depending on whether the pole would approach the positive or negative $\Im k_2$ axis, as shown in Fig.~\ref{fig:complexplane}. This is controlled by the sign of the $m_{22}$ parameter: if it is positive (negative) the resonance corresponds to a virtual 
(bound) state. From the figure one can also appreciate that poles on the II (IV) sheet are more likely to be bound (virtual) states,
as the sheet borders with the positive (negative) semiaxis.

We construct a training dataset of $10^5$ line shapes, generated by evaluating Eq.~\eqref{eq:intensity}
for intensity parameters uniformly sampled within a wide range of values (see Supplementary Material for details). 
We then obtain 65 intensity values by evaluating the line shape in the 
$\left[4.251,4.379\right]\gev$
invariant mass region in $2\mev$ bins. 
The intensity line shapes were convoluted with the experimental resolution, and 5\% Gaussian noise was added to the signal, to have statistical uncertainties that resemble the ones reported in Ref.~\cite{Aaij:2019vzc}. An additional validation set is generated to monitor the generalization performance of the model during training. To each line shape, we attach a label according to the bound/virtual nature and the Riemann sheet where the pole lays, \ie $b|2$, $b|4$, $v|2$, and $v|4$. Figure~\ref{fig:complexplane} shows examples of (noiseless) training line shapes for each class. 

\begin{figure}
        \includegraphics[width=\columnwidth]{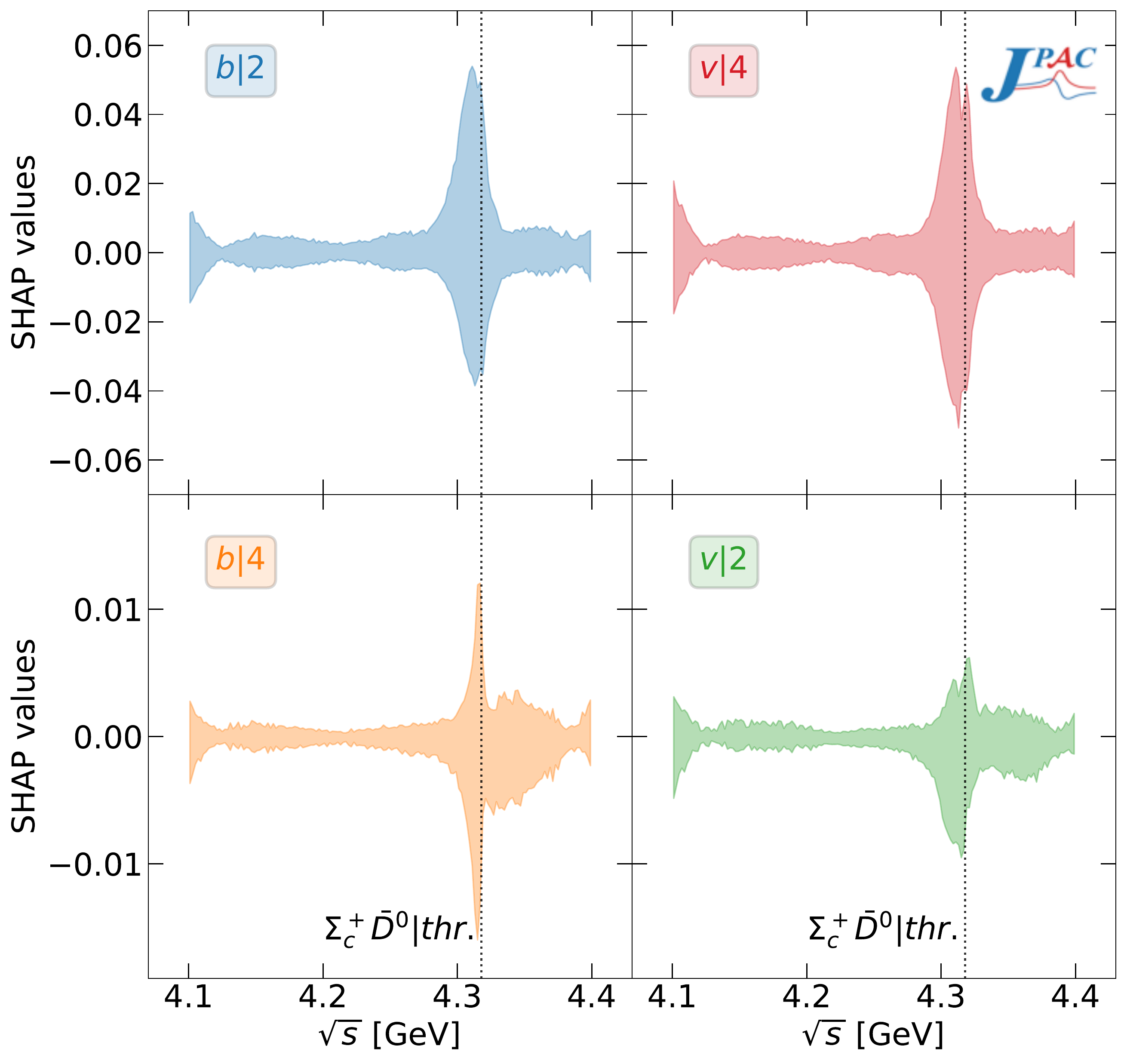} 
    \caption{Distribution of SHAP values as a function of the \jpsip invariant mass 
    for the four classes for the training data with a 5\% noise level.
    The shaded region corresponds
    to the 68\% confidence level. The 
    position of the \SigmaD\ threshold is highlighted. 
    It is apparent that such region has the largest impact.
    The amplification at the edges is a spurious border effect. }
    \label{fig:shaptraining}
\end{figure}

\mytitle{Training, validation, and invariant mass window}
We build a DNN
using the PyTorch framework~\cite{NEURIPS2019_9015}
that takes (noisy) line shapes as an input, and predicts the corresponding class $\{b|2, b|4, v|2, v|4\}$.
For optimization purposes, the line shapes are first rescaled between 0 and 1 as a normalization before we feed them into the network. The DNN consists of an input layer with as many nodes as there are energy bins, followed by two fully-connected hidden layers with 400 and 200 nodes respectively, and finally an output layer with four nodes that correspond to the four classes. 
After each hidden layer we set a dropout probability that randomly sets nodes to zero, to improve generalization performance. 
The DNN is trained using the Adam optimizer~\cite{DBLP}.
The details of the procedure are given in the Supplemental Material. 
We train the DNN for 100 
passes of the full training
dataset through the DNN,
{\it aka} epochs.
Figure~\ref{fig:confusedaccuracy} shows the training and validation sets accuracy for different levels of Gaussian noise, as well as the confusion matrix for the case of $5\%$ noise. 
This shows how the experimental uncertainty limits the accuracy, as expected. With this setup, the DNN learns the subtleties of the intensity line shapes associated with each one of the four different resonant pole structures. However, in order to obtain our final DNN classifier, we need to select the appropriate
invariant mass window around the \Pc\ signal where we allow the DNN to attribute importance. We introduce a systematic method based on SHapley Additive exPlanations (SHAP) values~\cite{NIPS2017_8a20a862}
to select a proper window. Using SHAP values, we can break down a prediction to show how each bin impacts classification. 
Therefore, we train a first DNN to  a wide range of invariant masses $\left[4.1,4.4\right]\gev$. 
A positive (negative) SHAP value indicates that a given data point is pushing the DNN classification in favor (against) a given class. 
Large absolute SHAP values imply a large impact of a given mass bin on the classification, as shown in Fig.~\ref{fig:shaptraining}.
The mass interval used here is the same as in~\cite{Fernandez-Ramirez:2019koa}.
This choice is confirmed by the
SHAP values analysis shown in
Fig.~\ref{fig:shaptraining}.
However, we checked that the results
are qualitatively unchanged even if a wider
window is selected.

\begin{figure}
        \includegraphics[width=\columnwidth]{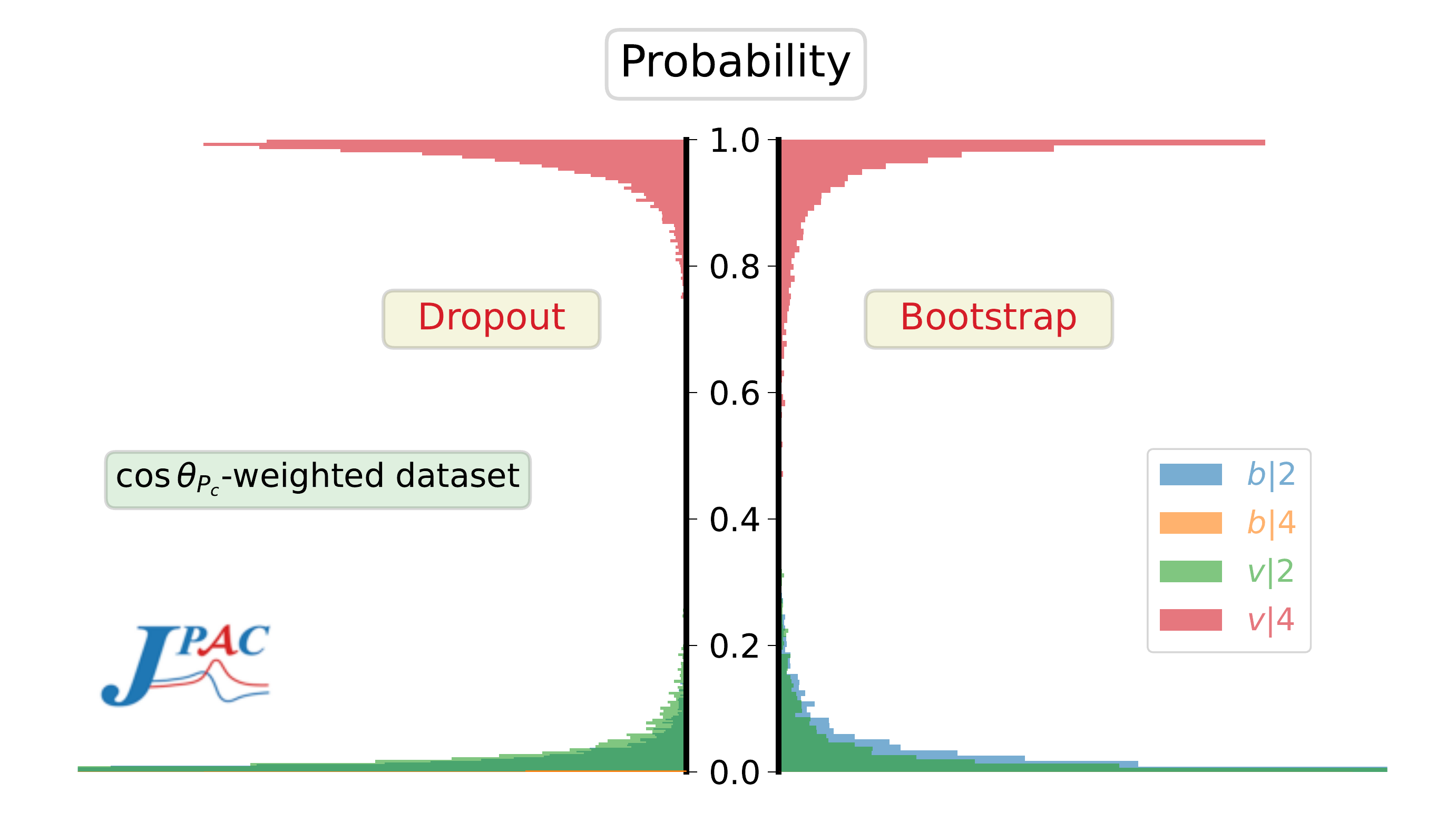} 
      \caption{Dropout and bootstrap classification probability densities for the predictions on $\cos \theta_{P_c}$-weighted LHCb dataset, for each of the four classes.
      The $b|4$ (barely visible) and $v|2$ classes concentrate their probability density near zero.
      The $x$ axis is cut for the purpose of visibility. The results for the other LHCb datasets are provided in the Supplemental Material.}\label{fig:dropoutvsbootstrap}
\end{figure}

\begin{table}
\caption{Softmax output probabilities~\cite{Goodfellow-et-al-2016}
for the three
experimental datasets by LHCb~\cite{Aaij:2019vzc}.
}
\begin{ruledtabular}
\begin{tabular}{l|cccc}
 & $b|2$ & $b|4$ & $v|2$ & $v|4$ \\
\hline
$\cos \theta_{P_c}$-weighted
 & $0.6\%$&$<0.01\%$&$1.1\%$& $98.3\%$\\
$m_{Kp}>1.9\gev$& $1.4\%$&$<0.1\%$&$1.6\%$& $97.0\%$\\
$m_{Kp}$ all &$5.4\%$&$<0.1\%$&$21.0\%$& $73.6\%$
\end{tabular}
\end{ruledtabular}\label{tab:results}
\end{table}

\begin{figure}
    \includegraphics[width=\columnwidth]{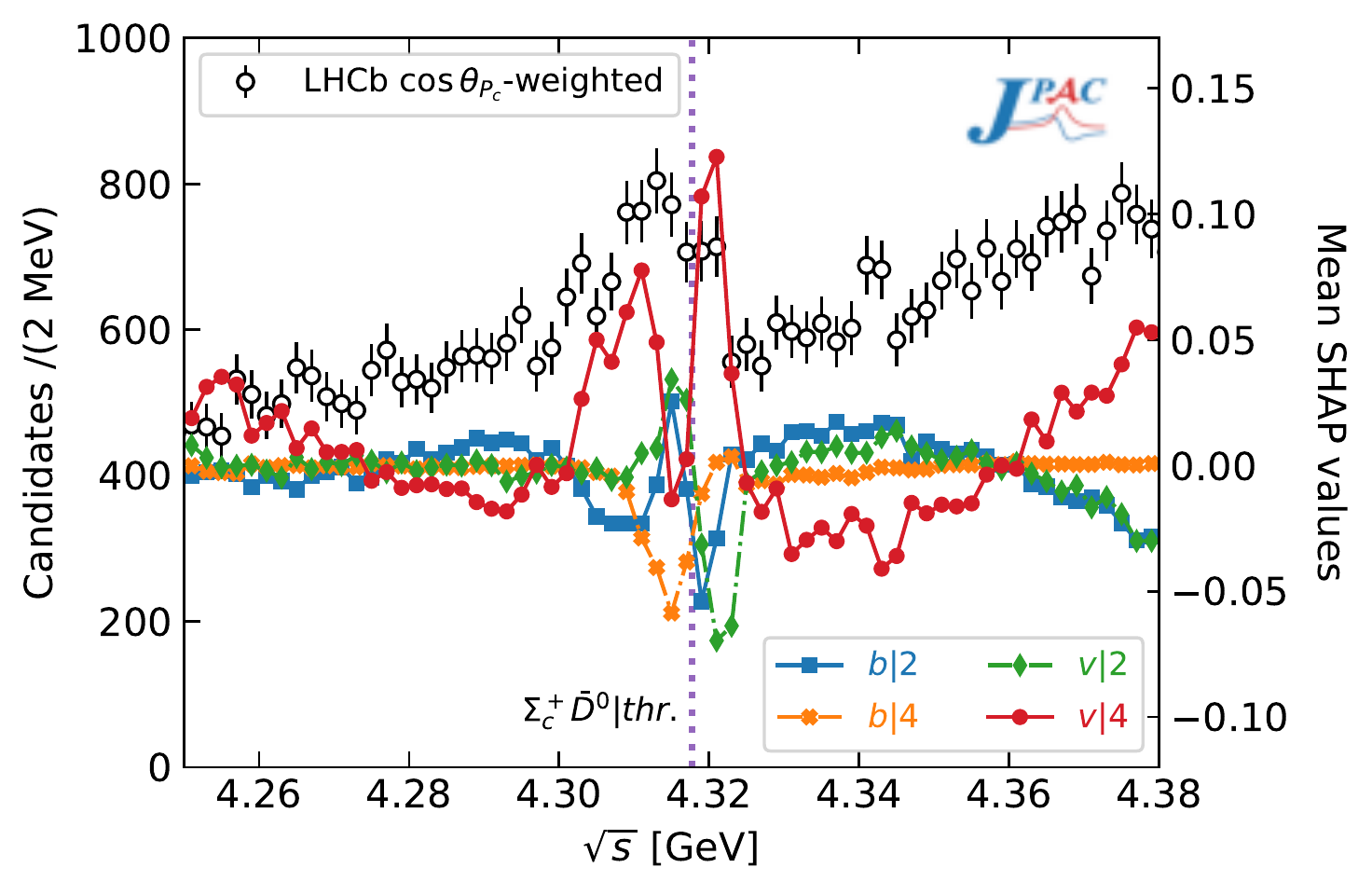} 
    \caption{$\cos \theta_{P_c}$-weighted LHCb
    data (left axis) and distribution of their mean 
    SHAP values (right axis) 
    as a function of the \jpsip invariant mass for the four classes. The results for the other LHCb datasets are provided in the Supplemental Material.}
    \label{fig:shaplhcb}
\end{figure}

\mytitle{Signal analysis}
We are now in a position to generate predictions on the nature of the pole on the actual experimental LHCb data. 
We pass the three datasets from~\cite{Aaij:2019vzc} through the DNN. We remind that one is the original $\Lambda^0_b \to K^- \jpsi\,p$ dataset, while two have sharp or smooth cuts that suppress the background from $\Lambda^*$ resonances.
The output probabilities
for each class are summarized in Table
\ref{tab:results}.
It is apparent that the virtual interpretation
is strongly favored, specifically the
$v|4$ class. To properly quantify the uncertainty of this prediction, we use two
Monte Carlo based methods: bootstrap~\cite{EfroTibs93,Landay:2016cjw}
and dropout~\cite{gal2016dropout}. Both methods aim at producing probability densities for the generated predictions on the LHCb data, as detailed in the Supplemental Material, and yield the same conclusions.
The probability densities of the four classes are shown in Fig.~\ref{fig:dropoutvsbootstrap}. 
Class $v|4$ is heavily preferred,
while $b|4$ is strongly rejected. 
Classes $b|2$ and $v|2$ attain low probabilities, in particular for the datasets with background rejection. Hence, we conclude that a virtual state with its pole placed on the IV Riemann sheet is the highly preferred interpretation of the \Pc\ signal.

The DNN classifier can provide further information on which invariant mass region contributes most strongly to this prediction, by repeating the SHAP analysis for the experimental data, as shown in Fig.~\ref{fig:shaplhcb}. It is apparent how the region close to threshold determines the DNN classification. Slightly above threshold, data favor the $v|4$ class, while rejecting the $v|2$ one.
Below threshold, the $v|4$ and $v|2$ classes are preferred to $b|2$, and $b|4$ is rejected.

\mytitle{Conclusions} 
We presented a proof of concept of how machine learning can be used to further our understanding
of exotic hadrons. 
We trained a neural network to learn the details of line shapes corresponding to different resonance interpretations, based on an effective range
expansion of the amplitude close to the relevant threshold. 
We apply this method to determine the nature of the \Pc\ signal seen by LHCb.
We determine the probability of each of the classes of interest, given the experimental uncertainties and resolution.  
 A DNN classifier significantly favors a virtual state interpretation, \ie generated by an attraction force not strong enough to form a bound state, thereby confirming the findings in Ref.~\cite{Fernandez-Ramirez:2019koa}. 
By adding SHAP value analyses, we study how each data point impacts the selection. This technique also allows to assess which set of physical variables (in this case the energy range) is relevant  to a specific hypothesis,  which in standard approaches is often a  heuristic guess. 
Our technique can be directly applied to other (non)exotic signals close to a threshold opening.

We foresee various followups. For example, 
one can reuse parts of DNN classifiers that generates line shape representations (i.e. parameters in the first layers), and reapply it to new data (so-called `transfer learning') to obtain general resonance classifiers across scattering channels, and predict which physics underlies other reaction data. 

\begin{acknowledgments}
This work was triggered within the
``Scattering theory and applications to hadrons" course
taught under the Indiana University Global Classroom initiative.
This work was supported by 
Polish Science Center (NCN) Grant No.~2018/29/B/ST2/02576,
UNAM-PAPIIT Grant No.~IN106921,
CONACYT Grant No.~A1-S-21389,
and U.S. Department of Energy Grants 
No.~DE-AC05-06OR23177, No.~DE-FG02-87ER40365, and DE-FG02-92ER40735.
AP has received funding from the European Union's Horizon 2020 research and innovation programme under the Marie Sk{\l}odowska-Curie grant agreement No.~754496.
CFR is supported by Spanish Ministerio de Educaci\'on y Formaci\'on Profesional under Grant No.~BG20/00133.
VM is a Serra H\'unter fellow and acknowledges support from the Spanish national Grants No.~PID2019–106080 GB-C21 and No.~PID2020-118758GB-I00.
\end{acknowledgments}
\bibliographystyle{apsrev4-2.bst}
\bibliography{quattro}
\newpage
\onecolumngrid
\section{Supplemental material}

\subsection{Theoretical intensity distribution} 
The intensity distribution is given by 
\begin{equation}
I(s)_\text{theo} = \rho(s) \left[ \abs{F(s)}^2   
 + B(s) 
\right],\label{1}
\end{equation} 
where 
\begin{equation}
\rho(s) =m_{\Lambda_b} p\, q \, ,
\end{equation}
is the phase space factor with 
 \begin{align}
 p &= \lambda^{1/2}(s,m^2_{\Lambda_b},m^2_K)/2m_{\Lambda_b}\, ,&
 q &= \lambda^{1/2}(s,m^2_p,m^2_\psi)/2\sqrt{s}\, ,
 \end{align}
 and $\lambda(x,y,z) = x^2 + y^2 + z^2 - 2xy - 2xz - 2yz$ is the K\"all\'en function.
We assume that the \Pc\ signal has well-defined spin and contributes to a single partial wave.
The background $B(s)$ is the smooth contribution
from all other partial waves, which adds incoherently
to the signal. Hence, we parametrize it as a first
degree polynomial $B(s)=b_0+b_1\, s$.
The amplitude $F(s)$ is the product of 
$P_1(s)$ and the $T_{11}(s)$ amplitude.
$P_1(s)$ is also a smooth function that provides
the production of $\jpsip\,K^-$
and absorbs the cross channel $\Lambda^*$ resonances projected onto the same partial wave as the \Pc. 
We also parametrize it 
as a first
degree polynomial $P_1(s)=p_0+p_1\, s$.
$T_{11}(s)$ describes 
the  $\jpsip\to\jpsip$ scattering.
Hence, $F(s)$ can be written
\begin{equation} 
F(s) = P_1(s)\, T_{11} (s),\quad \left(T^{-1}\right)_{ij} = m_{ij} - i k_i \,\delta_{ij}, \label{eq:caseAB}
\end{equation} 
where the $m_{ij}$ are constants
with $i,j =1,2$ corresponding to the $\jpsip$ 
and the \SigmaD\ channels, respectively. 
We define the momenta 
$k_1 = \sqrt{s - (m_{\psi} + m_{p})^2}$ 
and 
$k_2 = \sqrt{s - (m_{\Sigma^+_c} + m_{\bar{D}^0})^2}$. Unitarity would prescribe the replacement of $k_i$ by the two-body phase space.
We approximated it by a 
square root alone, which is consistent with the effective range expansion near threshold, and with the nonrelativistic limit.
The inclusion of the off-diagonal 
$P_2(s) T_{21}(s)$ term
does not change the analytic properties of the amplitude and its effects can be absorbed
into the parameters of $F(s)$.
We  stress that, since the \jpsip threshold is far away from the region of interest, this channel can effectively absorb all other channels with distant thresholds. Any contributions from further singularities are smooth in the region of interest and are effectively incorporated in the background parameters.
 For particle masses, we use the PDG values
 $m_{\Lambda_b}=5.61960$\gev,
 $m_{p}= 0.9382720813$\gev, 
 $m_{K}= 0.493677$\gev,
 $m_{\psi}= 3.0969$\gev,
 $m_{\bar{D}^0}=1.86483$\gev, and
 $m_{\Sigma_c^+} = 2.4529$\gev~\cite{pdg}.
The $\Sigma_c^+$ width has not been measured yet, but it is expected to be the same as its two isospin partners, $\Gamma_{\Sigma_c^0}1.83^{+0.11}_{-0.19}\mev$ and $1.89^{+0.09}_{-0.18}\mev$. We neglect it, as it is similar to the experimental resolution.

\subsection{Experimental energy resolution} 
The theoretical intensity is convoluted with the experimental resolution function
\begin{equation}
I(s)=\int_{m_\psi+ m_p}^{m_{\Lambda_b}-m_K} I(s')_\text{theo} \exp \left[-\frac{(\sqrt{s}-\sqrt{s'})^2}{2R^2(s)} \right] d\sqrt{s'} \Bigg/\int_{m_\psi+ m_p}^{m_{\Lambda_b}-m_K} \exp \left[-\frac{(\sqrt{s}-\sqrt{s'})^2}{2R^2(s)} \right] d\sqrt{s'} \, , \label{eqres}
\end{equation}
where 
\begin{equation}
R(s) = 2.71\mev - 6.56\times 10^{-6} \mev^{-1} \times \left(\sqrt{s} - 4567 \mev\right)^2
\end{equation}
is the experimental resolution
provided by LHCb in~\cite{Aaij:2019vzc}.

\subsection{Pole positions}
To assign each pole a respective Riemann sheet location we need to find the roots of the denominator of Eq.~\eqref{eq.ampl} which in momentum space is equivalent to solving the following algebraic equation for $q = -i k_2$,
\begin{equation}
    p_0 + p_1 \,q + p_2 \,q^2 + p_3 \,q^3 + q^4 = 0 
\end{equation}
with
\begin{subequations}
\begin{align}
    p_0 &= \left(s_1 - s_2\right) m_{22}^2 - \left(m_{12}^2 - m_{11}m_{22}\right)^2\\
    p_1 &= 2\left(s_1 - s_2\right)m_{22} + 2 m_{11}  \left(m_{12}^2 - m_{11} m_{22}\right)\\
    p_2 &= -m_{11}^2 + m_{22}^2 + s_1 - s_2\\
    p_3 &= 2m_{22}
\end{align}
\end{subequations}
where $s_1 = (m_\psi + m_p)^2$, $s_2 = (m_{\bar{D}^0} + m_{\Sigma_c^+})^2$, which yields the poles in $s$ solving $s = s_2 - q^2$.
The poles appear in conjugate pairs, each on a sheet identified by $(\eta_1,\eta_2)$ pair
\begin{align}
 \eta_1 & = \sign \text{Re} \left(\frac{m_{12}^2}{m_{22} + q} - m_{11}\right) &
\eta_2 &= \sign \text{Re} q  \, ,    
\end{align}
or by the customary naming scheme
\begin{subequations}
\begin{align}
\text{I sheet} &: (+,+)\, , &
\text{II sheet} &: (-,+)\, , \\
\text{III sheet} &: (-,-) \, , &
\text{IV sheet} &: (+,-) \, .
\end{align}
 \end{subequations}

\subsection{Dimensionality reduction and data visualization}
\begin{figure}
\includegraphics[width=\linewidth]{
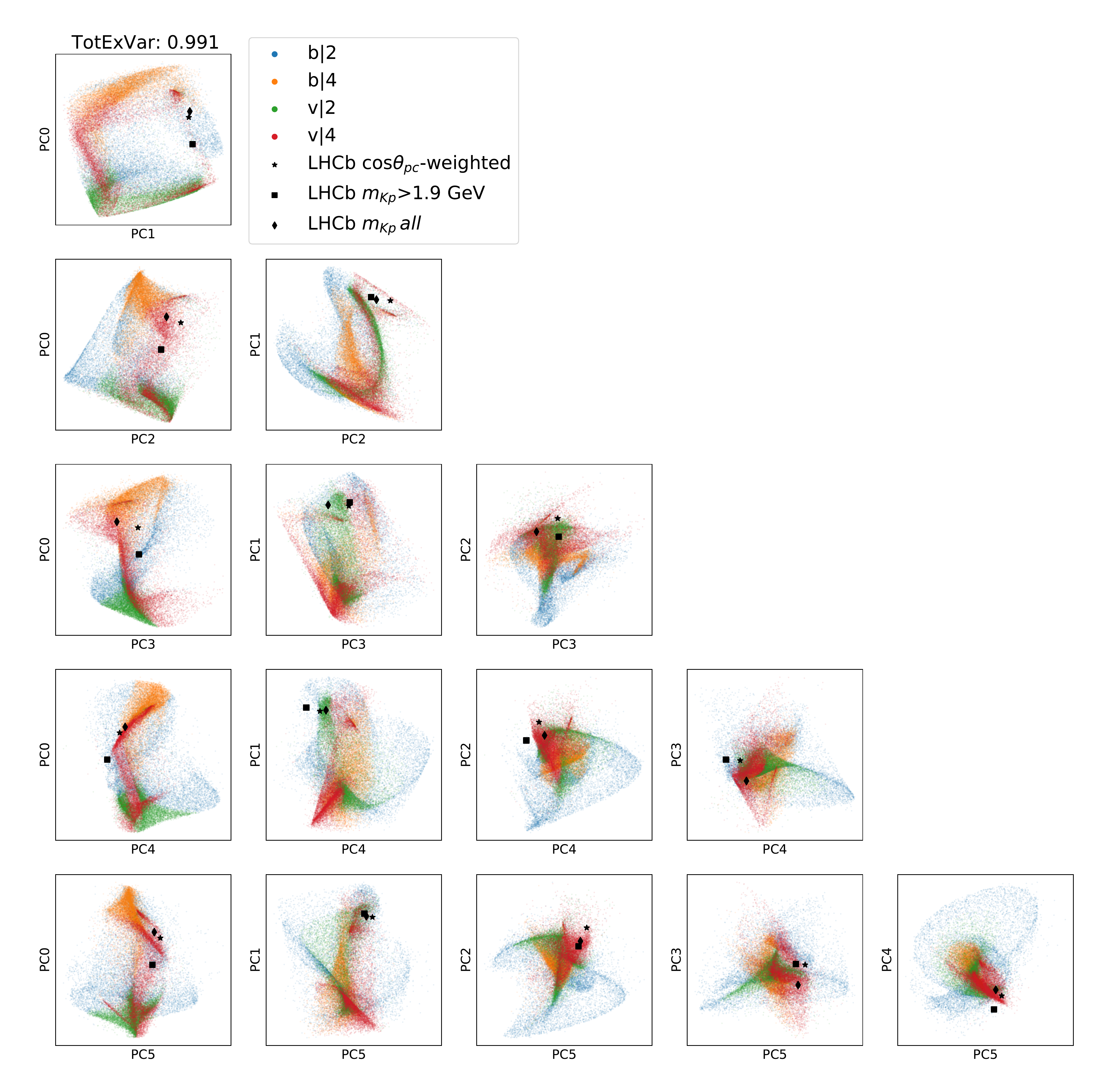}
\caption{Pairwise projections of training data samples onto the six principal components. Also shown are the projected three experimental LHCb datasets.}
\label{fig:pca_pairplot}
\end{figure}

\begin{figure}
\includegraphics[width=.6\linewidth]{
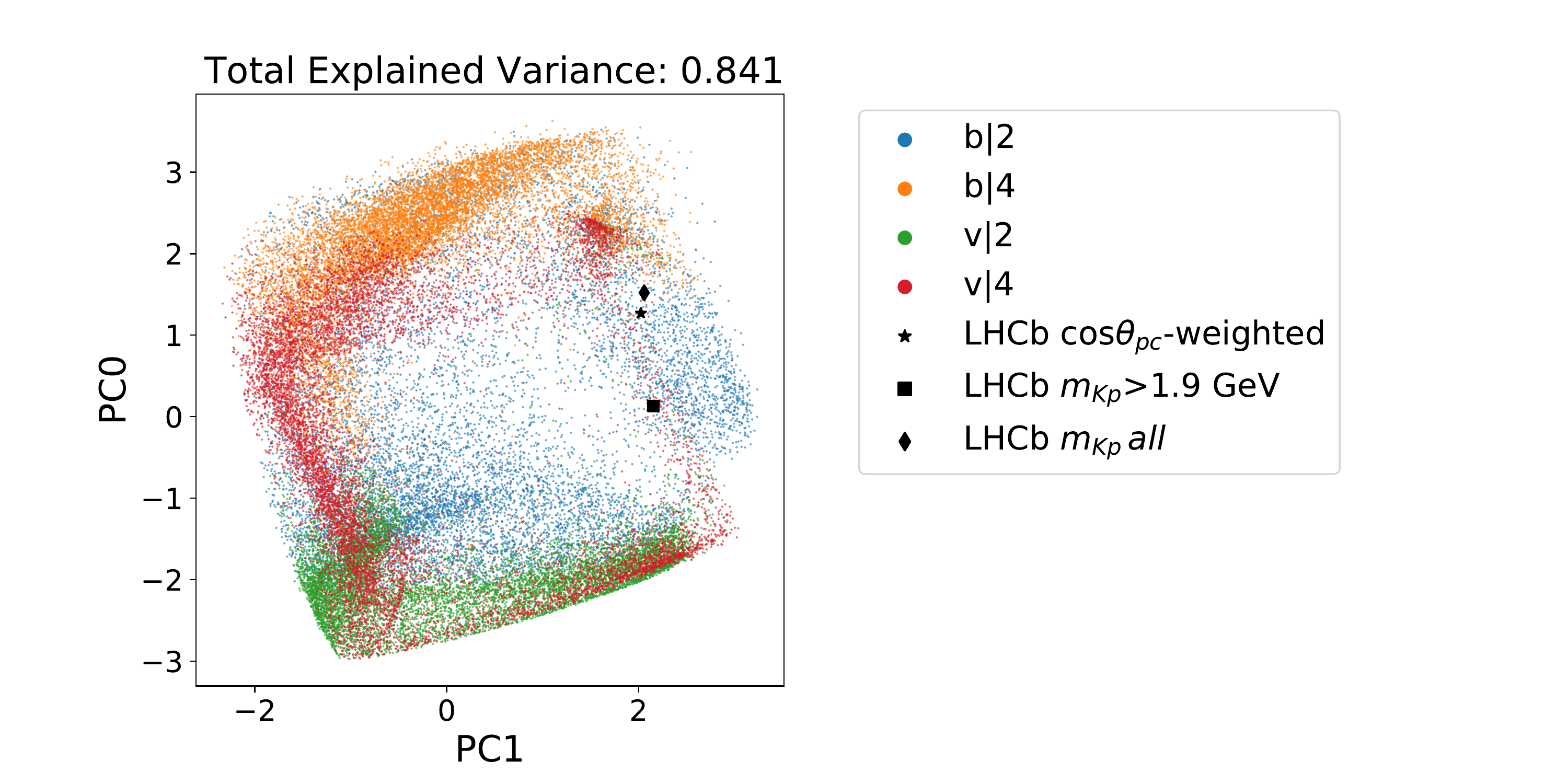}
\caption{Samples from the training data are projected onto a two-dimensional space using principal component analysis. The total explained variance in this two-dimensional projection is $0.841$.}
\label{fig:pca_pc0vspc1}
\end{figure}

In order to determine the location of the experimental \Pc\ mass spectrum, and whether it has support from the training data, we performed a Principal Component Analysis (PCA) to reduce the dimensionality~\cite{pearson1901planes}. PCA extracts the eigenvectors (Principal Components, PC) and eigenvalues of the training set's covariance matrix. The principal components are ordered by their normalized eigenvalues (which represent the explained variance) such that PC0 contains the largest fraction of dataset variance. Hereby, one can explain~99\% of the variance of the training set by retaining $6$ PCs. We show pairwise PC projections in Fig.~\ref{fig:pca_pairplot} of the training and experimental data set. It can be seen that the LHCb data points are well encompassed by the training data in the reduced PCA space, suggesting that the amplitudes developed can describe the experimental data well. An additional PCA with 2 PCs (with $84.1\%$ explained variance) is shown in Fig.~\ref{fig:pca_pc0vspc1}, resulting in the same conclusion. 

\subsection{Neural Network Architecture}
\begin{figure}
    \includegraphics[width=0.5\columnwidth]{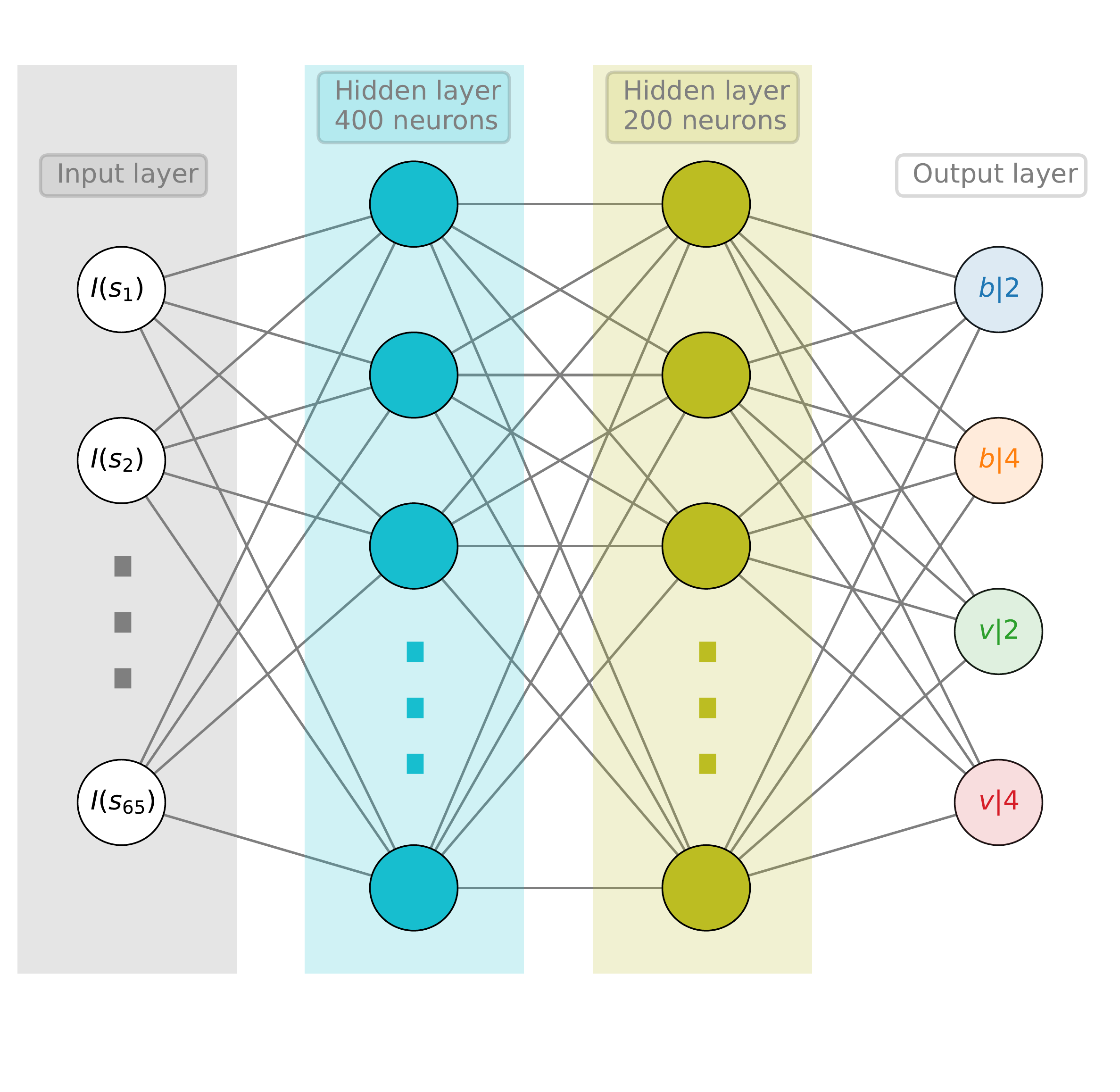} 
    \caption{Pictorial representation of the architecture of the DNN classifier with 65 input intensity points $I(s_n)$.}
    \label{fig:nn_architecture_sup}
\end{figure}

\begin{table}[htb]
\begin{ruledtabular}
    \caption{Detailed description of the architecture layers. All dense layers contain a bias term. We provide the shapes of the input and output tensors for a given batch size B, \ie for an input batch of dimension (B, 65). See main text for details.}
    \begin{tabular}{l|c|c}
        Layer  & Shape in & Shape out \\
        \hline \hline
        Input &   & (B, 65) \\
        Dense & (B, 65)   & (B, 400) \\
        Dropout(p=0.2) & (B, 400)   & (B, 400) \\
        ReLU    & (B, 400) & (B, 400) \\
        Dense & (B, 400)   & (B, 200) \\
        Dropout(p=0.5) & (B, 200)   & (B, 200) \\
        ReLU    & (B, 200) & (B, 200) \\
        Dense & (B, 200)   & (B, 4) \\
        Softmax    & (B, 4) & (B, 4)
    \end{tabular}
    \label{tab:architecture}
\end{ruledtabular}
\end{table}

A neural network is used in this study and is implemented in the PyTorch framework~\cite{NEURIPS2019_9015}. The architecture consists of two hidden layers with 400 and 200 nodes respectively. Rectified linear units (ReLU) are used as the activation function for these nodes [$\textrm{ReLU}(x) = \max(0,x)$]. After each of these hidden layers a dropout layer is included. This layer randomly masks nodes in a layer with a given probability, $p$. We use $p=0.2$ and $p=0.5$ for the first and second layer respectively. 
The final output layer consists of 4 nodes, each one representing a class. The softmax activation function is applied to the output of the final layer. This transformation
is a generalization of the logistic function to multiple dimensions. For a vector $x = [x_1, ..., x_C]^T$, with $C$ the number of classes, the softmax function
has the form~\cite{Goodfellow-et-al-2016} 
\begin{equation}
\textrm{softmax}(x_i)=\frac{e^{x_i}}{\sum_j{e^{x_j}}}\, .
\end{equation}
and allows the assignment of probabilities to each class.
The network is optimized by minimizing the multi-class cross entropy loss. The architecture is tabulated in Table~\ref{tab:architecture}. The Adam optimizer~\cite{DBLP} is used with a learning rate of 0.001. The network is trained for 100 epochs with a batch size of 1024. Although the architecture was initially obtained through an educated guess, it was verified using Bayesian hyperparameter optimization~\cite{NIPS2012_05311655}, which found an optimum close to the figures cited above.

\begin{figure}
    \includegraphics[width=0.5\columnwidth]{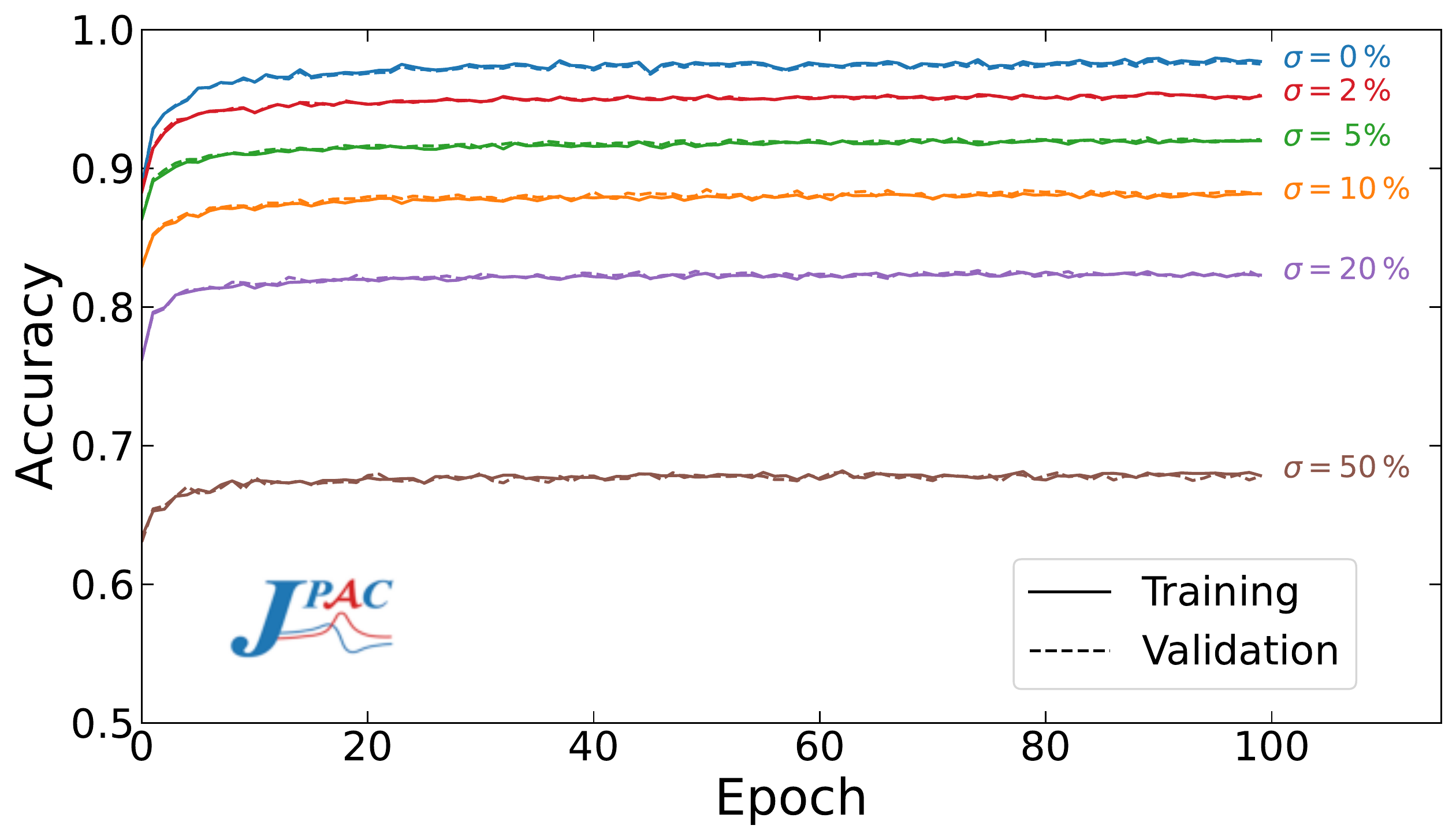} 
    \caption{Accuracy as a function of the epoch for the training and validation datasets in the $\left[4.251,4.379\right]\gev$ energy range. }
    \label{fig:epochs_sup}
\end{figure}

Figure~\ref{fig:epochs_sup} shows the evolution of the training and validation accuracy with the number of epochs. Multiple scenarios with varying noise levels are performed. Although the accuracy typically stabilizes after 40 epochs we let the DNN run up to 100 epochs. This result shows how the experimental uncertainty limits the accuracy, as expected.

\subsection{Uncertainty Quantification}

\begin{figure}
    \begin{tabular}{c}
    \includegraphics[width=0.48\columnwidth]{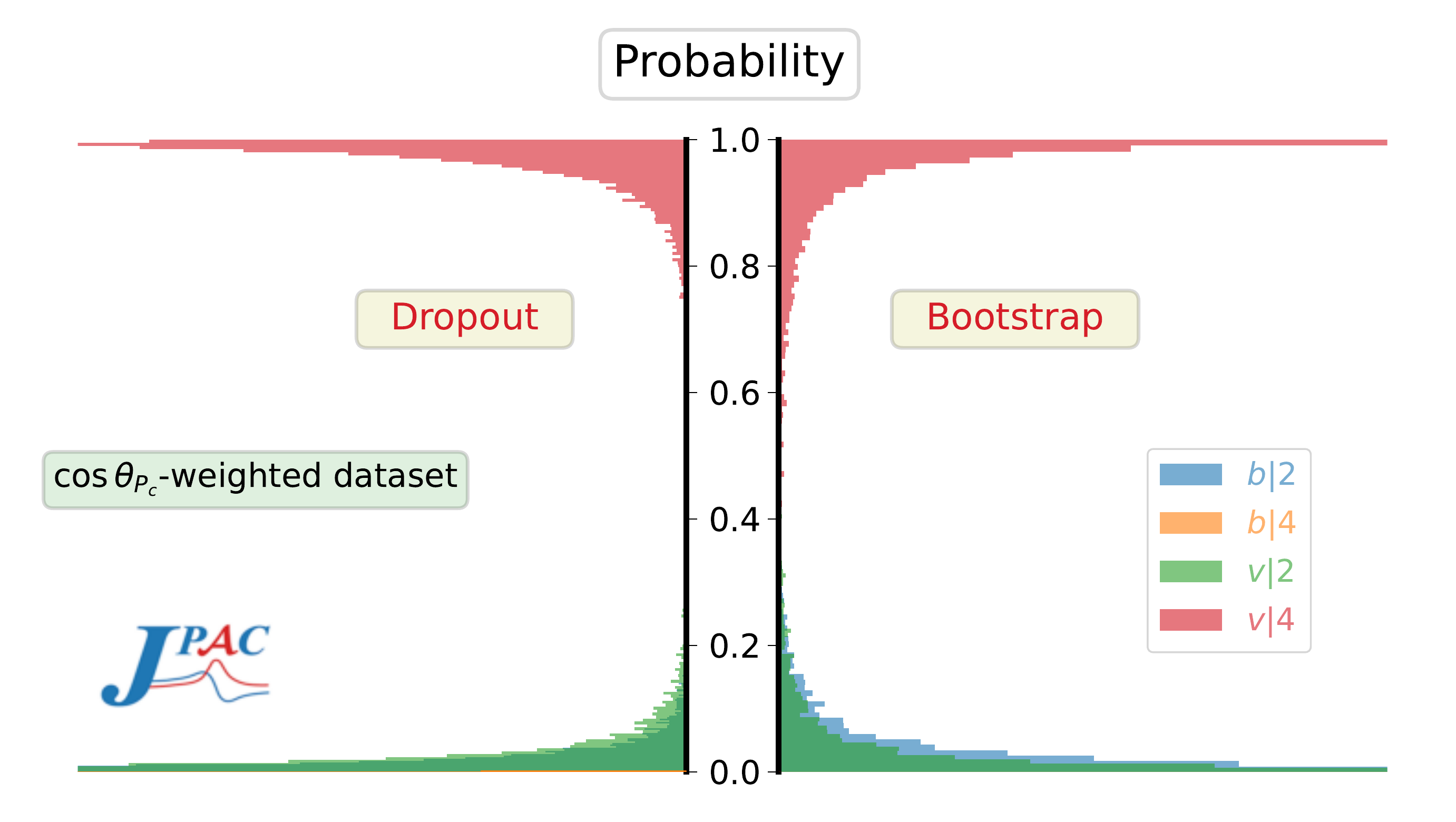} 
    \includegraphics[width=0.48\columnwidth]{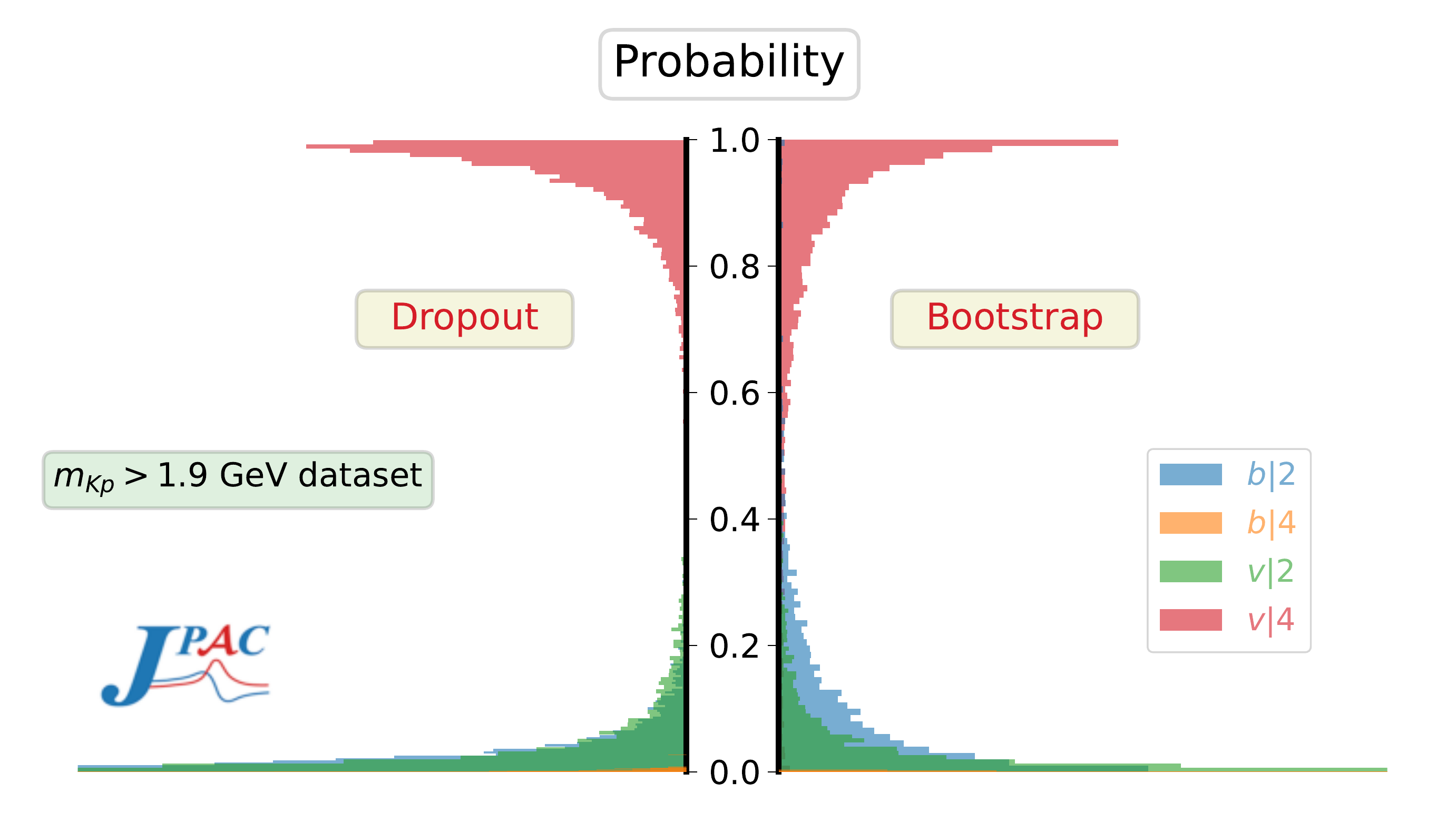} \\
    \includegraphics[width=0.48\columnwidth]{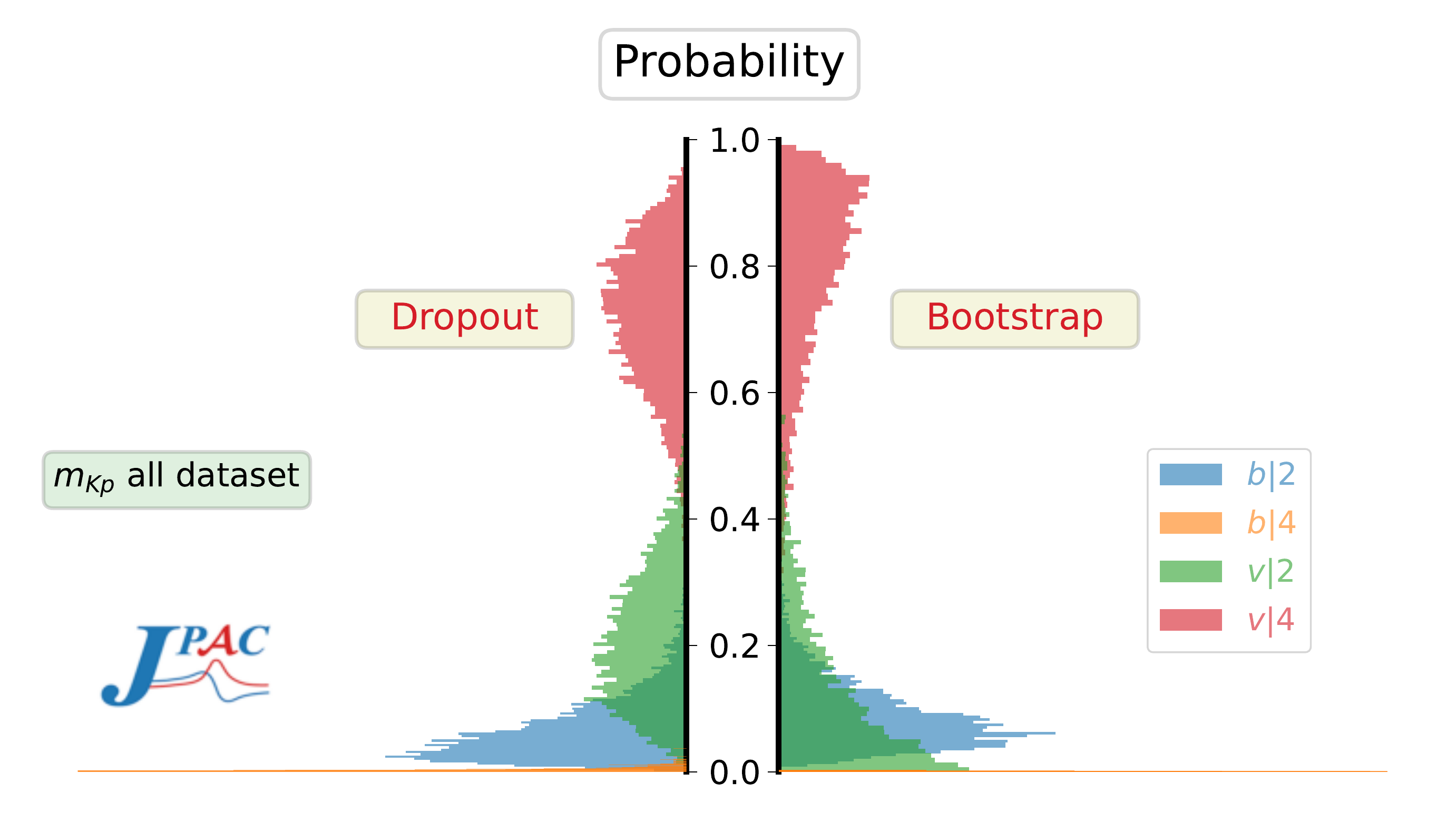} 
        \includegraphics[width=0.48\columnwidth]{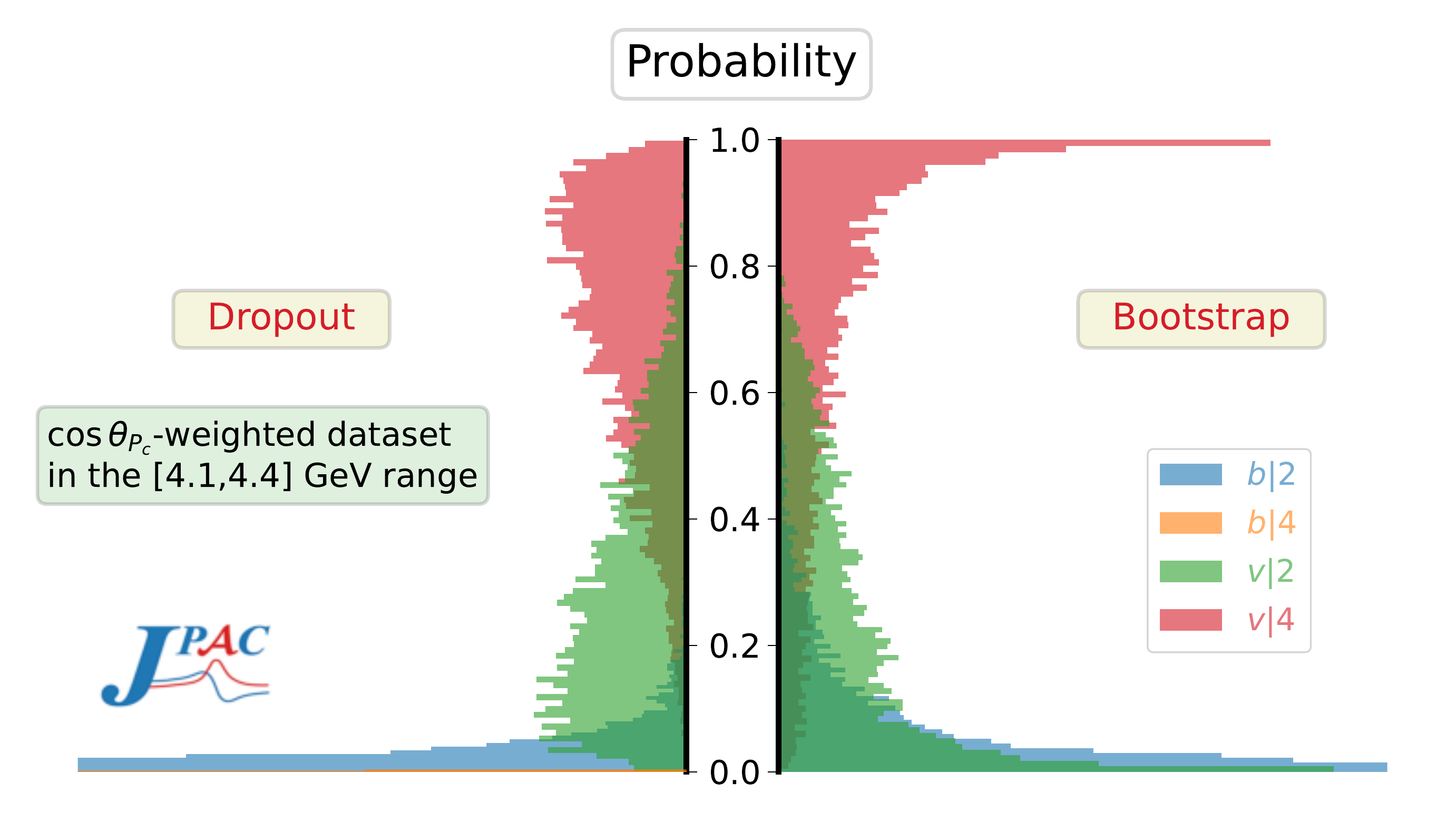} 
    \end{tabular}
    \caption{Dropout and bootstrap classification probability densities for the predictions on the three LHCb datasets in the $\left[4.251,4.379\right]\gev$ invariant mass range, and on the $\cos\theta_{P_c}$-weighted on the $\left[4.1,4.4\right]\gev$ range, for each of the four classes.
      The $x$ axes are equally cut for the purpose of visibility and comparison.}
    \label{fig:violin_sup}
\end{figure}

\begin{figure}
\includegraphics[width=0.7\columnwidth]{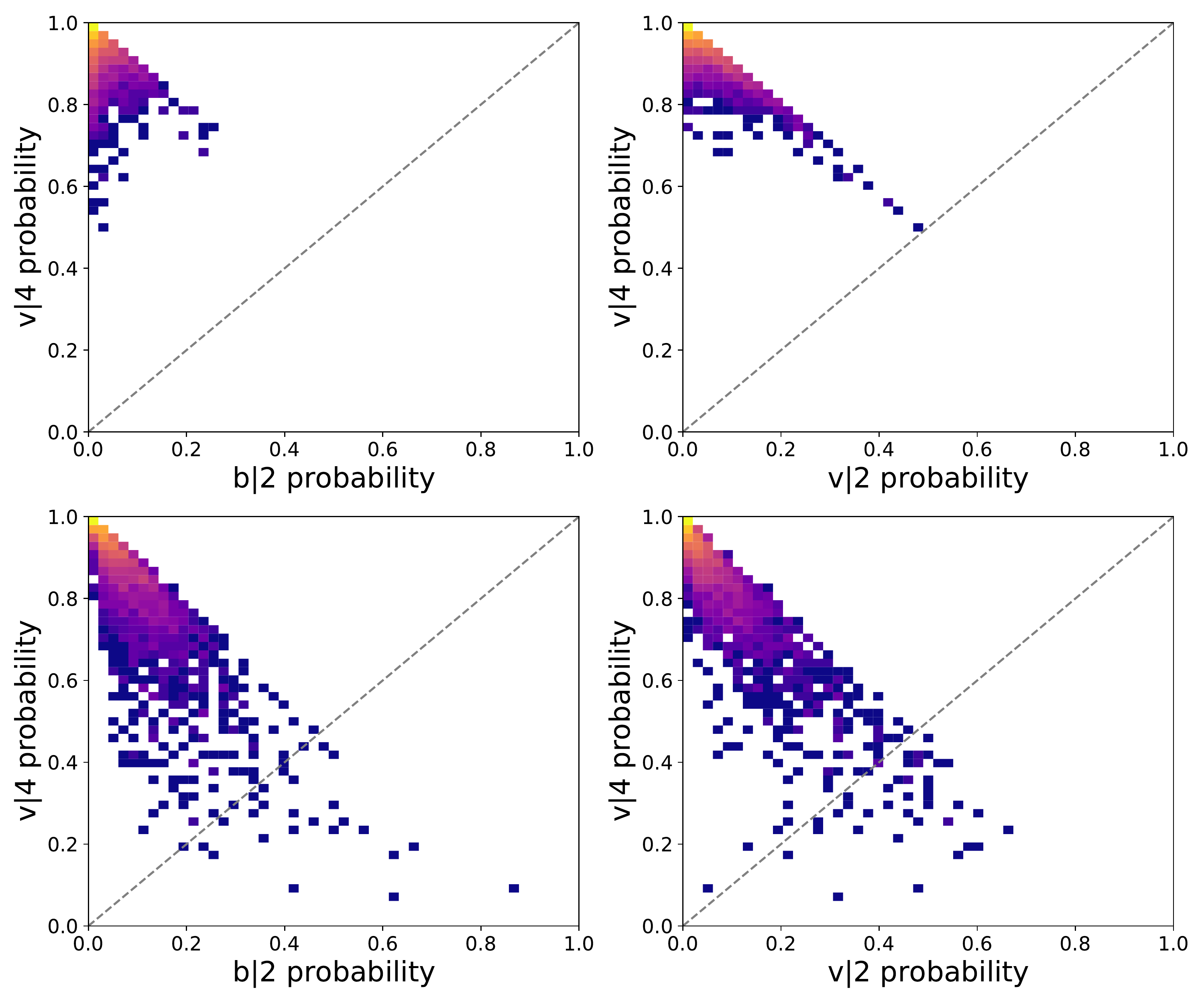}
\caption{Projections of the classification probability densities
using dropout (top row) and bootstrap (bottom row)
classification uncertainties with correlations for the
$\cos\theta_{P_c}$-weighted LHCb dataset in the $\left[4.251,4.379\right]\gev$ range .}\label{fig:corr_sup}
\end{figure}

The effect of the experimental uncertainties in the
classification needs to be properly quantified,
so we can answer to the question on
how strongly the prediction of our neural network depends on the instantiation of the statistical noise 
We use
two methods to extract uncertainties on the class predictions for the LHCb data. The bootstrap method uses 5000 samples of the LHCb data that are generated by sampling the LHCb data around its uncertainties, assuming the uncertainties are Gaussian. Then, the sampled data is passed through the network (without dropout). 
The second method (dropout, also known as Monte Carlo dropout) evaluates the model on the LHCb data (without error bars), but generates uncertainty distributions by sampling 5000 configurations for the dropout layers. Turning on dropout at evaluation time approximates Bayesian inference in a deep Gaussian process~\cite{gal2016dropout}. 
In Fig.~\ref{fig:violin_sup}
we show the resulting probability densities of the four classes and the three LHCb datasets in the $\left[4.251,4.379\right]\gev$ range.
For completeness, we also show the result for the 
$\cos\theta_{P_c}$-weighted dataset
for the wide $\left[4.1,4.4\right] \gev$ range.
This range covers datapoints far away from the region of interest and it is apparent how the virtual interpretation of the signal is still favored. Figure~\ref{fig:corr_sup} shows the correlation of the probability densities obtained with both methods
for the $\cos\theta_{P_c}$-weighted dataset in the $\left[4.251,4.379\right]\gev$ range.

\subsection{SHAP Values}
\begin{figure}
    \includegraphics[width=0.48\columnwidth]{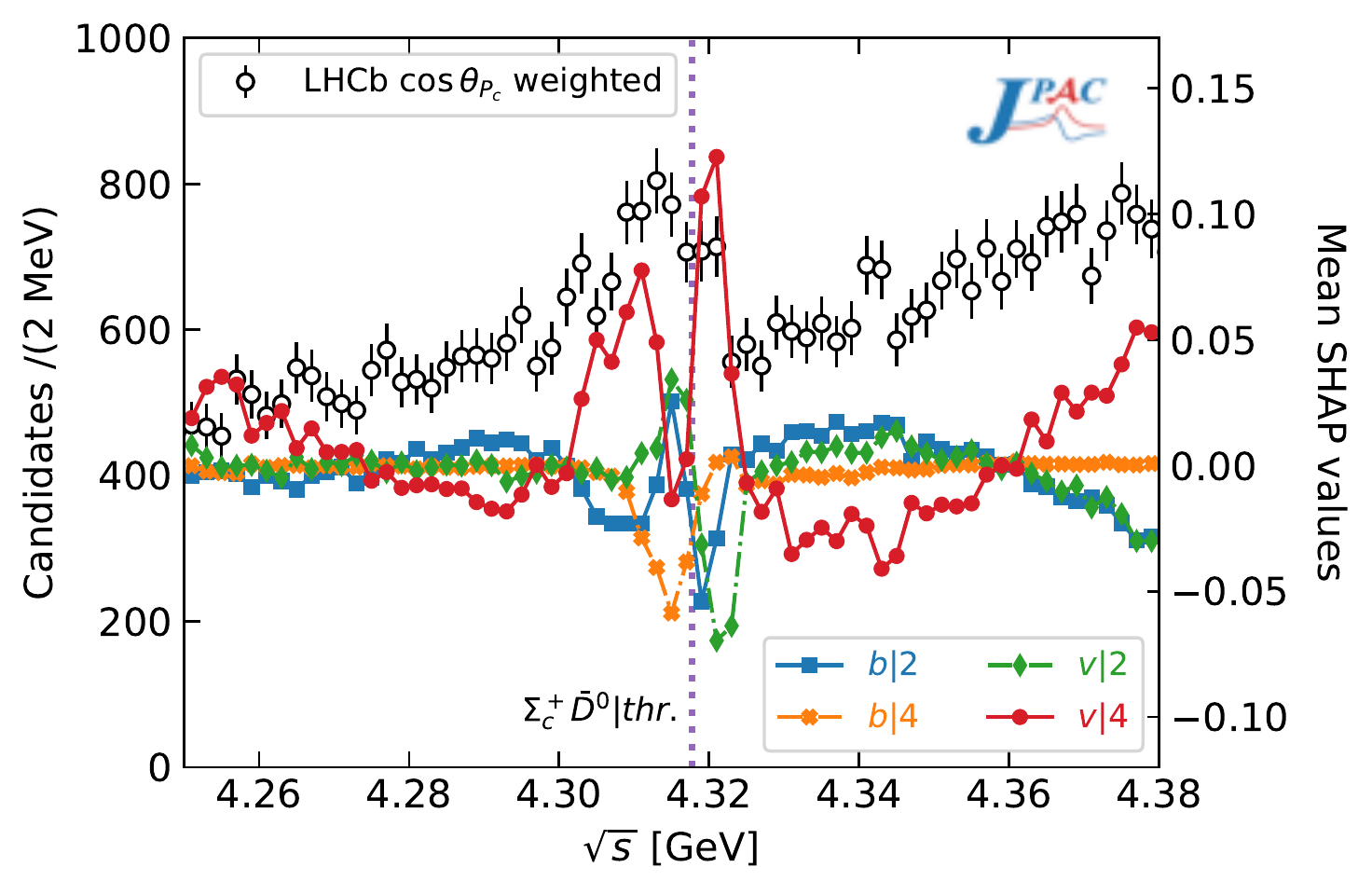} 
    \includegraphics[width=0.48\columnwidth]{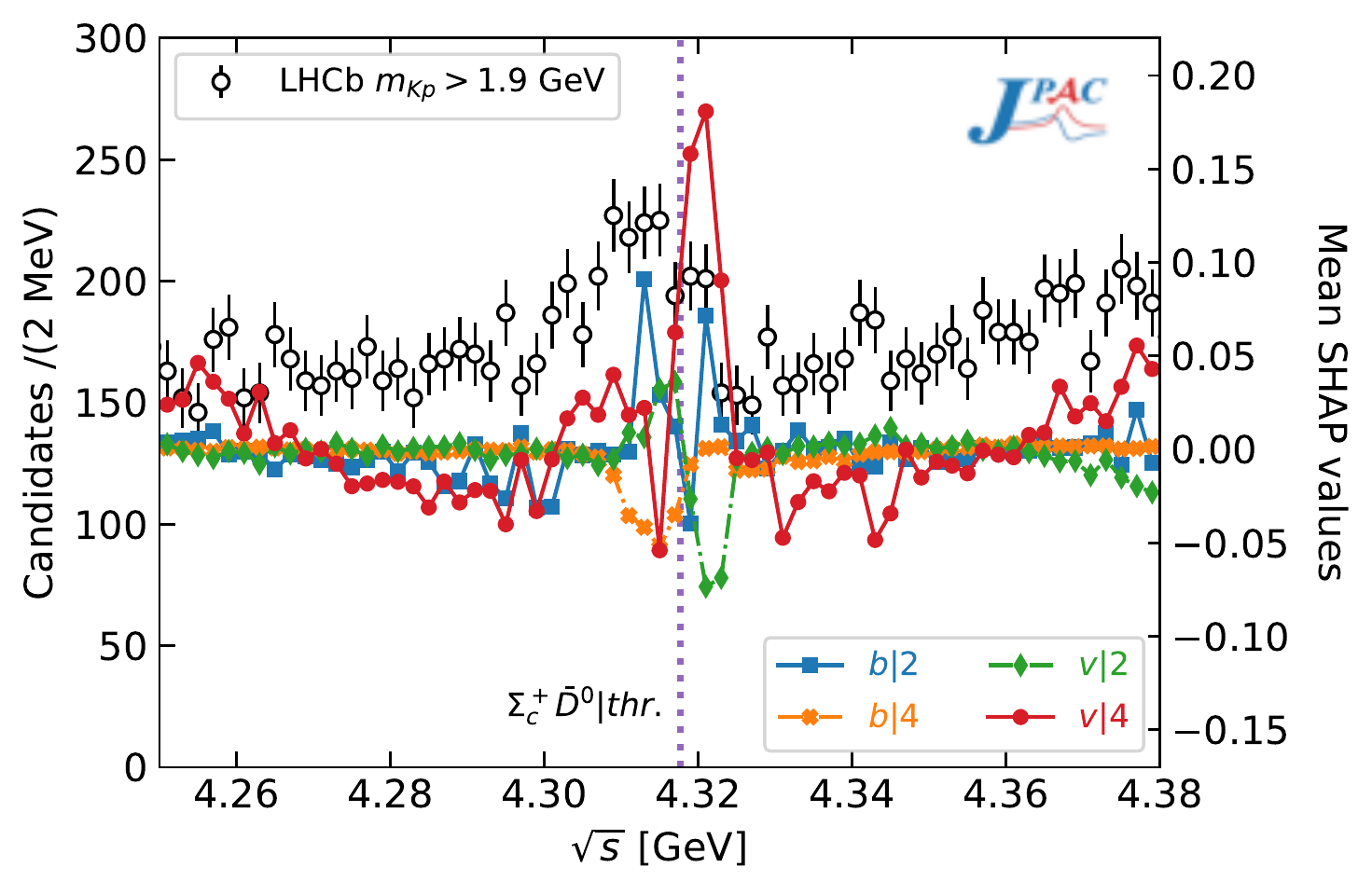} \\
    \includegraphics[width=0.48\columnwidth]{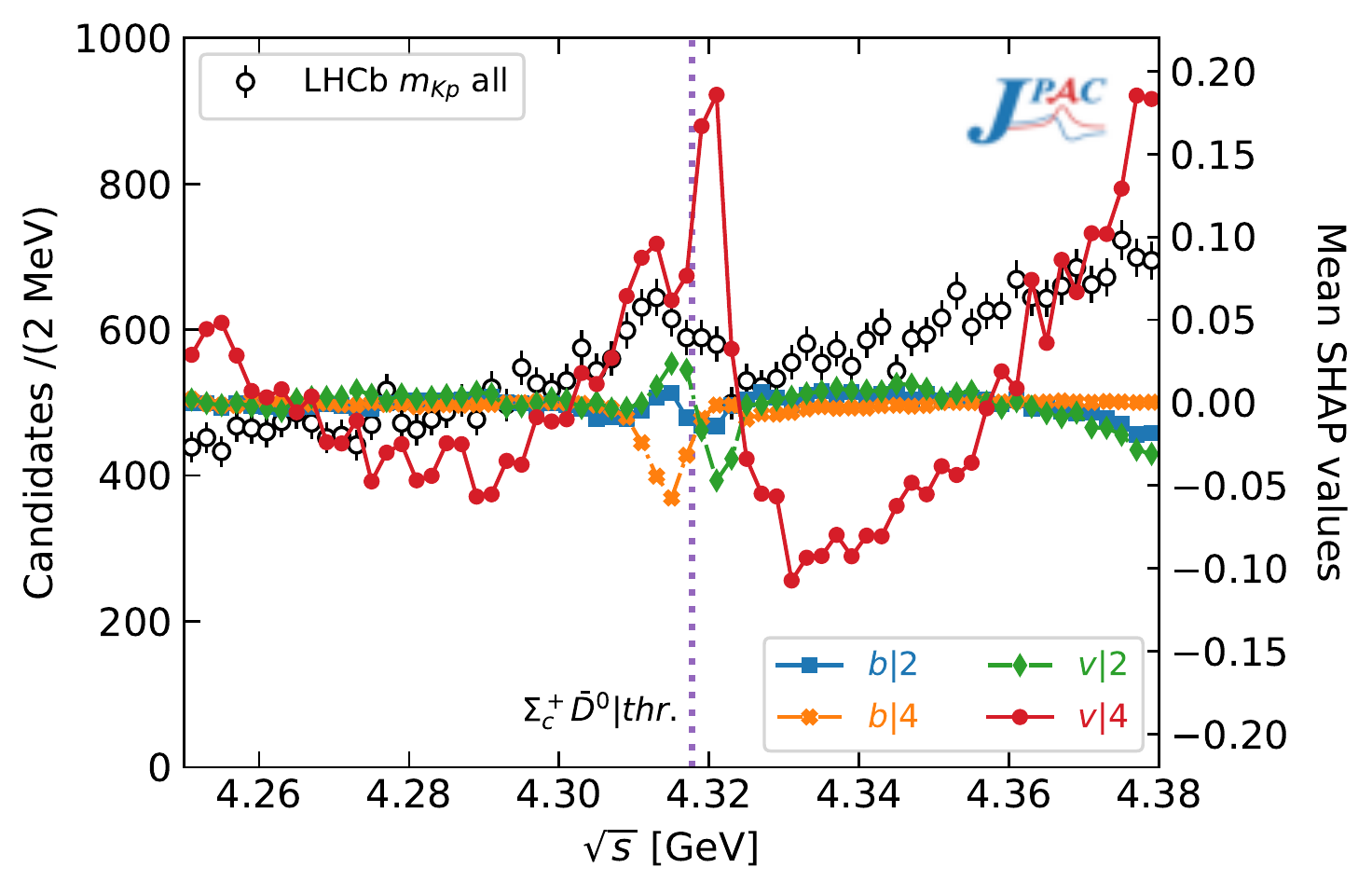} 
    \caption{LHCb
    data (left axis) and distribution of their mean 
    SHAP values (right axis) 
    as a function of the \jpsip invariant mass for the three LHCb datasets and the four classes.}
    \label{fig:shap_sup}
\end{figure}

SHapley Additive exPlanations (SHAP) values~\cite{NIPS2017_8a20a862} allow
to assess the impact of each datapoint in the
prediction the DNN makes compared to a certain baseline.
First we use them to validate the restricted invariant mass region used in the final analysis of the signal 
(See Fig. 3 of the manuscript),
later we used them to quantify which class is favored by
each experimental datapoint.
Figure~\ref{fig:shap_sup} shows the results for the
three datasets reported by LHCb in~\cite{Aaij:2019vzc}. 
In particular, we use the 
DeepExplainer implementation, which extracts approximate SHAP values~\cite{NIPS2017_8a20a862}).
The method is model agnostic and identifies the input contributions that lead to a certain prediction in a "fair" way.
More specifically, the SHAP values represent the average marginal contribution of a feature (in our case, the datapoint at a certain invariant mass) over all feature coalitions (\ie the whole set of considered datapoints). The marginal contribution is the difference in prediction with and without the feature. Each coalition represents a subset of the total feature dimensionality. DeepExplainer requires a background dataset to integrate over in order to extract approximate SHAP values. These SHAP values represent the difference in prediction between the background model output and the current model. A background dataset of 3000 spectra is taken and the SHAP values are computed for 1000 spectra. 

\end{document}